\documentclass{PoS}
\usepackage{journal_macros}
\usepackage{amsmath}
\usepackage{color, xcolor}
\usepackage[utf8]{inputenc}

\title{Galactic cosmic ray propagation models using Picard}

\ShortTitle{Galactic cosmic ray propagation models using Picard}

\author{\speaker{R. Kissmann}$^a$, O. Reimer$^a$, and A.~W. Strong$^b$\\
\llap{$^a$}Institute for Astro- and Particle Physics, Innsbruck University\\
\llap{$^b$}Max-Planck-Institut für extraterrestrische Physik\\
E-mail: \email{ralf.kissmann@uibk.ac.at}, \email{olaf.reimer@uibk.ac.at}, \email{aws@mpe.mpg.de}}

\abstract{We present results obtained from our newly developed Galactic cosmic-ray transport code \textsc{Picard}, that solves the cosmic-ray transport
equation. This code allows for the computation of cosmic-ray spectra
and the resulting gamma-ray emission. Relying on contemporary
numerical solvers allows for efficient computation of models with deca-parsec resolution. \textsc{Picard} can handle locally anisotropic spatial
diffusion acknowledging a full diffusion tensor. We used this framework
to investigate the transition from axisymmetric to spiral-arm cosmic-ray source distributions. Wherever possible we compare model
predictions with constraining observables in cosmic-ray astrophysics.}

\FullConference{The 34th International Cosmic Ray Conference \\
		 30 July- 6 August, 2015\\
		 The Hague, The Netherlands}

\begin{document}

\section{Introduction}
The \textsc{Picard} code for the numerical simulation of Galactic cosmic-ray transport was introduced in \cite{Kissmann2014APh55_37}. Similar to other software packages like \textsc{Dragon} (see \cite{EvoliEtAl2008JCAP10_18}) and \textsc{Galprop} (see, e.g., \cite{StrongMoskalenko1998APJ509_212, MoskalenkoStrong1998ApJ493_694}) \textsc{Picard} solves the cosmic-ray transport equation:
\begin{align}
  \label{EqTransport}
  \frac{\partial \psi_i}{\partial t}
  =&
  q(\vec{r}, p)
  +
  \nabla\cdot(\mathcal{D} \nabla \psi_i - \vec{v} \psi_i)
  +
  \frac{\partial}{\partial p} p^2 D_{pp}
  \frac{\partial}{\partial p} \frac{1}{p^2} \psi_i
  \\
  \nonumber
  &-
  \frac{\partial}{\partial p}
  \left\{
  \dot p \psi_i - \frac{p}{3} (\nabla \cdot\vec{v})\psi_i
  \right\}
  - 
  \frac{1}{\tau_f} \psi_i
  -
  \frac{1}{\tau_r} \psi_i
\end{align}
where $q$ is the parametrisation of the cosmic-ray source density, $\mathcal{D}$ is the spatial diffusion tensor, $\vec{v}$ is the spatial advection velocity, $D_{pp}$ represents diffusive re-acceleration, the subsequent term represents normal and adiabatic energy changes, and $\tau_f$ and $\tau_r$ are the time scales for fragmentation and radiative decay. The transport equation describes the time-evolution of the distribution function $\psi_i$ of cosmic ray species $i$. The interaction between the different cosmic-ray species is realised via the source and loss terms.

Like with \textsc{Galprop} and \textsc{Dragon}, \textsc{Picard} solves the transport equation for the cosmic-ray distribution function throughout the whole Galaxy to allow for a computation of secondary emission channels as gamma rays and neutrinos. This has to be seen in contrast to semi-analytical solution models (e.g., \textsc{Usine} -- see \cite{PutzeEtAl2009AnA497_991, PutzeEtAl2010AnA516_A66}), which allow for a faster computation of the cosmic-ray spectrum at Earth but do not contain a sufficiently detailed model of the Galaxy to model gamma-ray emission.


In \textsc{Picard} the focus is on finding steady-state solutions to the transport equation for a variety of aspects in Galactic cosmic ray transport.
For this we use several contemporary numerical solvers particularly suitable for an efficient solution of the cosmic-ray transport equation using distributed memory computers. By this we achieve high spatial resolution with the potential to investigate small-scale substructures of the Galaxy in corresponding modelling efforts. In the following we discuss the features of the \textsc{Picard} code. Additionally, we illustrate the capabilities of \textsc{Picard} by some example models using spiral-arm cosmic-ray source distributions.

\section{The \textsc{Picard} code}
Here, we will only give a brief overview of the features of the \textsc{Picard} code and the implementation of the solver; a more detailed description of the numerical solver of the \textsc{Picard} code can be found in \cite{Kissmann2014APh55_37}. A particular property of the \textsc{Picard} solver is the focus on spatially three-dimensional problems.

\subsection{Numerical solver}
In \textsc{Picard} we featured different solvers for time-dependent and for steady-state problems. In most other codes the latter kind of problems is also handled with the same solution scheme as is used for time-dependent problems just by integrating until the changes are below a given level (see, e.g., \cite{StrongEtAl2007ARNPS57_285}).

In \textsc{Picard} we solve the steady-state problem directly by inverting the matrix equation resulting from the discretisation of the steady-state form of the transport equation. The corresponding matrix is a block-diagonal matrix. There is a range of contemporary algorithms for the solution of the corresponding system of algebraic equations available. Here, we rely on the multigrid method (see \cite{TrottenbergEtAlBook2001}) and the BICGStab solver (see, e.g., \cite{sleijpen1993bicgstab}). These solvers are implemented in a form that they can also be used on distributed-memory machines, thus allowing high-spatial-resolution simulations. These solvers are of iterative nature, where as a stopping criterion the residual of the discrete transport equation has to become smaller than a given limit. It turns out that when using a time-integration scheme instead ensuring the convergence of the algorithm needs more care by the user (see \cite{KissmannEtAl2015APh70_39}).

For time-dependent problems a steady-state solution can be computed as an initial condition, thus, allowing to use the simulation for the time-variability aspect of the problem instead of waiting for the convergence to some initial state. \textsc{Picard} also includes different numerical schemes to handle time-dependent problems, but a solution via an explicit scheme is usually recommended since resolving the small dynamical time scales for the highest energy particles is necessary anyhow.

The entire code runs either as a single-processor or an MPI-parallel version, thus, allowing efficient use of large-scale computing structure. The numerical solver treats one single species at a time. Thus, the realisation of the interdependence in the nuclear reaction network needs to be taken into account. This is done by repeated solution of the transport equation for each species. The efficiency of this highly depends on the order in which the different species are handled, where we found a very efficient approach in \textsc{Picard} as is discussed in \cite{KissmannEtAl2015APh70_39}.

\subsection{Propagation physics}
In solving the transport equation the \textsc{Picard} code uses a similar prescription for the transport parameters as in \textsc{Galprop} or \textsc{Dragon}.
Since spatially three-dimensional problems are handled very efficiently using \textsc{Picard}, we focus on including small-scale features of the Galactic model that are problematic to be handled in two-dimensional or low-resolution three-dimensional models. A particular example is the investigation of the effect of a distribution of the cosmic-ray sources related to the spiral-arm structure of our Galaxy that will be discussed in the subsequent section.

Apart from that, \textsc{Picard} allows to use a general form of the spatial diffusion tensor, where also off-diagonal elements are acknowledged. These need to be taken into account whenever the predominant direction of spatial diffusion does not coincide with the directions of the numerical grid. Effects of such an anisotropic diffusion tensor have already been studied previously in the context of spiral-arm cosmic-ray source distributions (see \cite{EffenbergerEtAl2012AnA547A120}), where the authors showed a significant impact of such an anisotropic diffusion tensor. Apart from that \cite{EvoliEtAl2012PhRvL108_1102} used spatially variable diffusion as an alternative explanation to the cosmic-ray gradient problem.

\textsc{Picard} is built in a highly modular fashion that allows to include more detailed description of the propagation physics with relative ease. In the following we will focus on the example of a spiral-arm cosmic-ray source distribution.

\section{Results for spiral-arm cosmic-ray source distribution models}
Supernova remnants as one of the most prominent cosmic-ray sources are thought to be correlated with the higher star-formation rate in the vicinity of the Galactic spiral arms (see \cite{Reynolds2008ARAnA46_89, BerezhkoVoelk1997APh7_183, EllisonEtAl2004AnA413_189, AmatoBlasi2006MNRAS371_1251}). But also other cosmic-ray sources like pulsars are mostly rather young objects and should, thus, occur near the sites of star formation. Therefore, we investigated the impact of a cosmic-ray source distribution concentrated along the Galactic spiral arms. 

\begin{figure}
	\centering
	\includegraphics[width=0.72\textwidth]{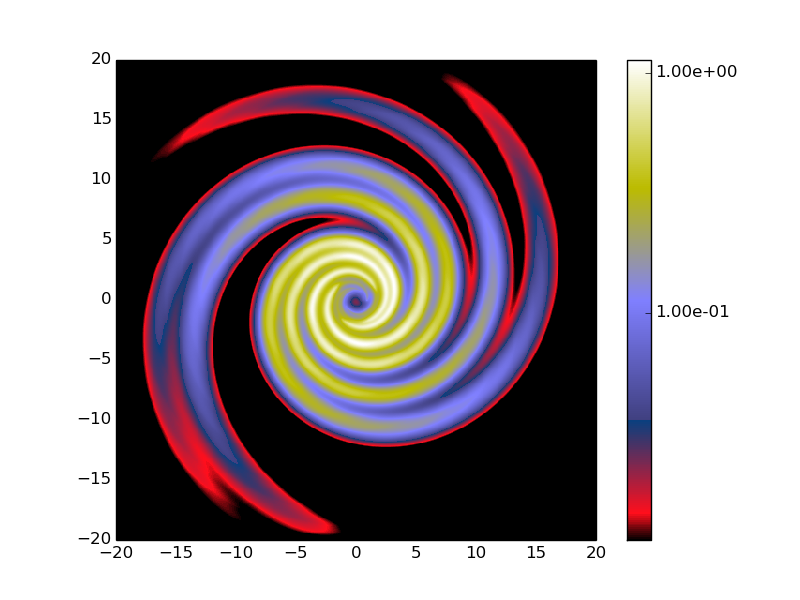}
	\caption{\label{FigElectronsExtreme}Distribution of 1\,TeV electrons in the Galactic plain for a plain-diffusion propagation model including a four-arm logarithmic-spiral source distribution.}
\end{figure}

The impact of spiral-arm source distributions on the different cosmic ray observables has been studied recently by several groups including \cite{BenyaminEtAl2014ApJ782_34, EffenbergerEtAl2012AnA547A120, GaggeroEtAl2013PhRvL111_021102, KoppEtAl2014NewA30_32} and \cite{WernerEtAl2015APh64_18}.
Results obtained for such a study for the highest resolution so far done with \textsc{Picard} are shown in Fig. \ref{FigElectronsExtreme}. The corresponding simulation used a spatial resolution of $\sim$75\,pc in each direction for a plain diffusion model with parameters adapted from \cite{StrongEtAl2010ApJ722L_58S}. In Fig. \ref{FigElectronsExtreme} we show the distribution of $\sim$1\,TeV electrons in the Galactic plane. There the confinement of high-energy electrons to their source regions becomes apparent. For electrons this confinement becomes even more pronounced at highest energies. This also results in a change of the electron spectrum, when comparing the flux at an on-arm to an inter-arm position. For nuclei this effect becomes much weaker due to the weaker energy losses. 

Due to the high memory demand the spatial resolution was reduced to $\sim$150\,pc, when the effects of the nuclear reaction network were taken into account. The effect of the source localisation becomes particularly apparent, when investigating the secondary to primary ratios as are discussed in, e.g. \cite{BenyaminEtAl2014ApJ782_34} and \cite{KissmannEtAl2015APh70_39}.

In the latter of these papers \textsc{Picard} was applied to study the impact on the secondary to primary ratios for different spiral-arm source distributions. We investigated several of these source distributions in order to account for the variety among present-day Milky Way models. Apart from that the spiral-arm structure appears different when observed via different tracers.
 Not only do the different tracers relate to different position within the spiral arms (see \cite{Vallee2014ApJS215_1}), but some tracers only show two spiral arms instead of four as inferred from other tracers (see, e.g., the discussion in \cite{Steiman-CameronEtAl2010ApJ722_1460}).

Therefore, we investigated a range of cosmic-ray source models that should cover the possible extreme configurations for logarithmic spiral-arm models: we used a logarithmic two-arm and a logarithmic four-arm model, designated as the Dame- and the Steiman-model, respectively, and additionally a parametrisation based on a model for the distribution of free electrons in the Galaxy by \cite{CordesLazio2002astro_ph_7156} (referred to as the NE2001-model). Thus, we cover the different extreme cases for Galactic spiral-arm models, where, e.g, the NE2001-model features broader spiral arms than the others and also includes a local spiral arm segment not taken into account in the other models.

The transport equation for the different models was solved using reference transport-parameter sets adapted from \cite{AckermannEtAl2012ApJ750_3} (for details see Table 1 in \cite{KissmannEtAl2015APh70_39}).

The resulting cosmic ray fluxes for the different nuclei at Earth inferred from the different models are still in approximate agreement with the observed cosmic-ray data. This result could have been expected from the observation by \cite{WernerEtAl2015APh64_18} that the spectral slopes of cosmic-ray nuclei are only weakly affected by the source distribution. For cosmic-ray nuclei the dominant effect is the variation of the total flux. Only for the extreme case of the two-arm model do the differences relative to an axisymmetric source distribution become visible in the spectra at Earth.

\begin{figure}\setlength{\unitlength}{0.00045\textwidth}
	\begin{minipage}{0.5\textwidth}
    \begin{picture}(1100,845)(-100,-20)
        \put(-50,420){\rotatebox{90}{$y$}}%
	    \includegraphics[width=1000\unitlength]{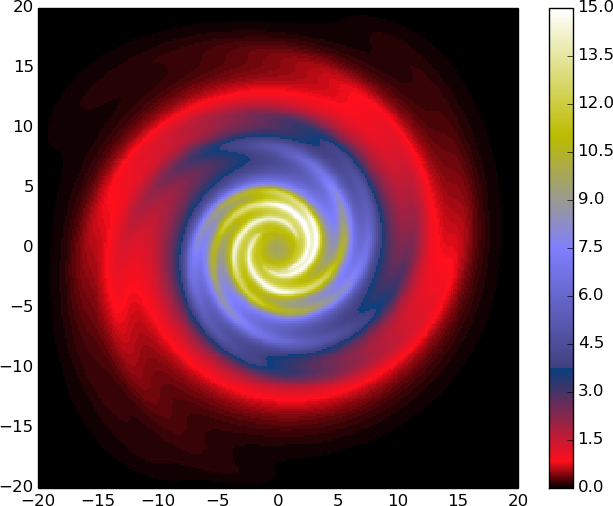}
    \end{picture}
    ~\\
    \begin{picture}(970,292)(-130,-100)
        \put(-50,90){\rotatebox{90}{$z$}}%
        \put(416,-50){$x$}%
        \includegraphics[width=840\unitlength]{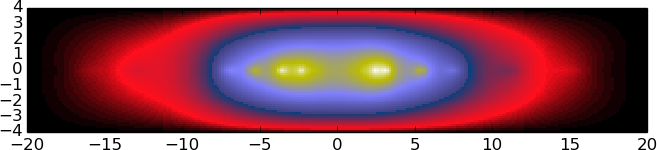}
    \end{picture}
    \end{minipage}
    \begin{minipage}{0.5\textwidth}
    \begin{picture}(1100,845)(-100,-20)
        \put(-50,420){\rotatebox{90}{$y$}}%
        \includegraphics[width=1000\unitlength]{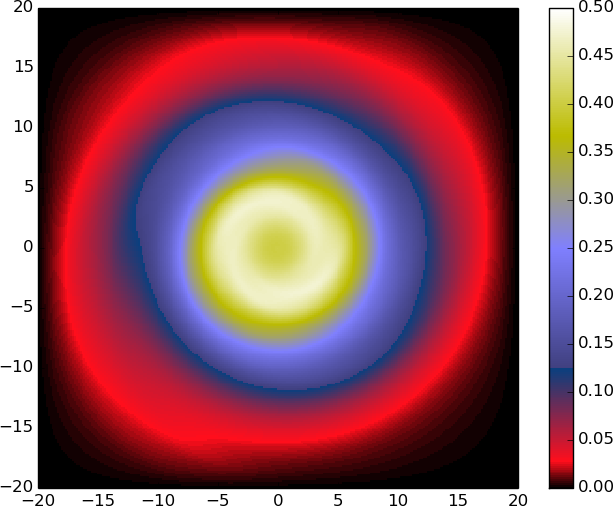}
    \end{picture}
        ~\\
    \begin{picture}(970,292)(-130,-100)
        \put(-50,90){\rotatebox{90}{$z$}}%
        \put(416,-50){$x$}%
        \includegraphics[width=840\unitlength]{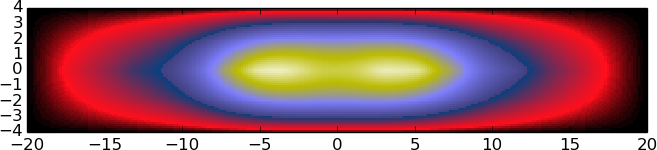}
    \end{picture}
    \end{minipage}
    \caption{\label{FigSpiralArmDist}Flux of $\sim$10\,GeV Galactic cosmic rays in the Galactic plane (top) and in the $x-z$-plane (bottom). Results are shown for $^{12}$C (left) and for $^{10}$B (right) for a four-arm source distribution. Spatial scales are given in kpc. The Earth is located at $x=8.5$\,kpc, $y=z=0$\,kpc.}
\end{figure}

The effect of the localised source distribution becomes apparent when comparing primary and secondary cosmic rays. In Fig. \ref{FigSpiralArmDist} we illustrate this difference by comparing the resulting distributions of $\sim$10\,GeV $^{12}$C and $^{10}$B in the Galaxy. As expected, $^{12}$C -- as a typical primary -- follows the source distribution more closely than a secondary like $^{10}$B. In \cite{KissmannEtAl2015APh70_39} we discuss accordingly, how this depends on the source-distribution. The effect is strongest in the two-arm source distribution, but is also strong in the Steiman-model. In the NE2001-model the effect is much less pronounced since the source distribution is much smoother in that case.

Resulting from this, we observe that secondary to primary ratios like B/C show strong local variation when the model uses a localised source distribution. Regarding a fit to the observed secondary to primary ratios, the distance from the observer to the nearest spiral arm becomes degenerate with some of the other transport parameters like the strength of re-acceleration and of spatial diffusion. Of course, the position of the spiral arms is by no means arbitrary, but it is subject to some variation in the different models (due to different tracers or due to a possible time-lag between the current position of the spiral arms and the peak of the cosmic-ray source distribution (see \cite{Shaviv2003NewA8_39})).

Using the set of propagation parameters from \cite{AckermannEtAl2012ApJ750_3} only the Steiman-model yields results in accordance with the cosmic-ray data at Earth within the uncertainties of the spiral-arm models. A moderate alteration of the other transport parameters, however, also allows for a good fit to the observations for the NE2001-model (as shown in \cite{KissmannEtAl2015APh70_39}). For the Dame two-arm model, however, we were not able to find an acceptable fit. Thus, such a two-arm model in the current form seems to be inconsistent with the cosmic-ray data.

This discussion shows that in order to assess the quality of a cosmic-ray transport parameter set, we need additional data. Including the axisymmetric model we have three models with rather different spatial distributions of the cosmic rays that are consistent with cosmic-ray observations at Earth. Thus, a measure of the spatial distribution of the cosmic rays is necessary to evaluate the different models. The most convenient measure of this kind is the Galactic diffuse gamma-ray emission (see \cite{AckermannEtAl2012ApJ750_3}), which is currently investigated for the different models.

Two measures, more directly related to cosmic rays albeit not as sensitive as the gamma-ray emission, are the cosmic-ray gradient and the long-term variation of the cosmic-ray flux at Earth. When neglecting the long-term changes of heliospheric modulation the NE2001-model does not recover a recurrent imprint of the spiral arms on the long-term cosmic-ray flux as is expected according to \cite{Shaviv2003NewA8_39}. The same model, however, is the only one, which is consistent with the finding that the cosmic-ray flux in the recent past is higher than the long-term average (see \cite{Shaviv2003NewA8_39}). This is caused by the presence of the local spiral-arm segment in the NE2001-model, which is not explicitly part of the other spiral-arm models.

This local spiral-arm segment has another important impact on the cosmic-ray distribution: we found that the NE2001-model, in contrast to all other models investigated in this study, is consistent with the observed cosmic-ray anisotropy in the 1\,TeV range. Actually, the anisotropy for the NE2001-model is about an order of magnitude smaller than for the other models. While there might also be other explanations for the observed cosmic-ray anisotropy (see, e.g., \cite{EvoliEtAl2012PhRvL108_1102}), this finding shows that the local spiral-arm segment included in the NE2001-model has significant impact on the distribution of cosmic rays near Earth. Together with the missing imprint of the spiral-arm structure on the long-term cosmic-ray flux variation in the NE2001-model this indicates that a consideration of the local spiral-arm segment within the Steiman-model might constitute the most sensible test for the cosmic-ray source distribution currently at hand. Apart from that it clearly shows that a more detailed description of the source distribution of cosmic rays -- and also of other transport parameters -- will be essential when aiming at explaining all observations related to Galactic cosmic rays.

\section{Conclusion \& Outlook}
Here, we discussed the features of the recently introduced \textsc{Picard} code for the numerical investigation of the propagation of Galactic cosmic rays. Apart from detailing some of the features of \textsc{Picard}, we illustrated the capabilities of the code by the example of a study of the impact of spiral-arm cosmic-ray source distributions on the propagation of cosmic rays. This discussion showed that a localised source distribution of the cosmic rays has an important impact on different cosmic-ray observables. However, a possible fit to the cosmic-ray spectra at Earth shows that assessing the quality of a model should be done in conjunction with some measure of the spatial distribution of cosmic rays. Furthermore, this study indicates that a more detailed model of our Milky Way in the context of cosmic-ray propagation is possible and also necessary to investigate the ever improving wealth of data of the different observables connected to cosmic rays.

\bibliography{$HOME/LaTeX/Bibliographies/GalacticCR,$HOME/LaTeX/Bibliographies/pubrk.bib,$HOME/LaTeX/Bibliographies/numerics.bib}



\end{document}